\documentclass{article}

\usepackage{PRIMEarxiv}

\usepackage[utf8]{inputenc} 
\usepackage[T1]{fontenc}    
\usepackage{hyperref}       
\usepackage{url}            
\usepackage{booktabs}       
\usepackage{amsfonts}       
\usepackage{nicefrac}       
\usepackage{microtype}      
\usepackage{lipsum}
\usepackage{fancyhdr}       
\usepackage{graphicx}       
\usepackage{authblk}        
\usepackage{tikz}           
\usepackage{natbib}
\setcitestyle{authoryear} 

\usepackage{amsmath}

\graphicspath{{media/}}     

\title{Addressing researcher degrees of freedom through minP adjustment}

\author[1,5]{\large{\textbf{Maximilian M Mandl}\thanks{
\textbf{e-mail: mmandl@ibe.med.uni-muenchen.de}} }}
\author[1,2]{\textbf{Andrea S Becker-Pennrich}}
\author[2,3]{\textbf{Ludwig C Hinske}}
\author[4]{\textbf{Sabine Hoffmann}}
\author[1,5]{\textbf{Anne-Laure Boulesteix}}
\affil[1]{Institute for Medical Information Processing, Biometry, and Epidemiology, Faculty of Medicine, LMU Munich, Germany}
\affil[2]{Department of Anaesthesiology, LMU University Hospital, LMU Munich, Germany}
\affil[3]{Institute for Digital Medicine, University Hospital of Augsburg, University of Augsburg, Germany}
\affil[4]{Department of Statistics, LMU Munich, Germany}
\affil[5]{Munich Center for Machine Learning (MCML), Germany}

\begin{document}
\maketitle

\begin{abstract}
When different researchers study the same research question using the same dataset they may obtain different and potentially even conflicting results. This is because there is often substantial flexibility in researchers' analytical choices, an issue also referred to as \lq\lq researcher degrees of freedom''. Combined with selective reporting of the smallest p-value or largest effect, researcher degrees of freedom may lead to an increased rate of false positive and overoptimistic results.  
In this paper, we address this issue by formalizing the multiplicity of analysis strategies as a multiple testing problem. As the test statistics of different analysis strategies are usually highly dependent, a naive approach such as the Bonferroni correction is inappropriate because it leads to an unacceptable loss of power. Instead, we propose using the  \lq\lq minP'' adjustment method, which takes potential test dependencies into account and approximates the underlying null distribution of the minimal p-value through a permutation-based procedure. This procedure is known to achieve more power than simpler approaches while ensuring a weak control of the family-wise error rate. 
We illustrate our approach for addressing researcher degrees of freedom by applying it to a study on the impact of perioperative $paO_2$ on post-operative complications after neurosurgery. A total of 48 analysis strategies are considered and adjusted using the minP procedure. This approach allows to selectively report the result of the analysis strategy yielding the most convincing evidence, while controlling the type 1 error---and thus the risk of publishing false positive results that may not be replicable. 
\end{abstract}


\newpage
\section{Introduction}
\label{sec:intro}
In recent years, the scientific community has become increasingly aware that there is a high analytical variability when analysing empirical data, i.e. there are plenty of sensible ways to analyse the same dataset for addressing a given research question, and they may yield (substantially) different results \citep{gelman2014statistical,silberzahn2018many}. 
If combined with selective reporting, this variability may lead to an increased rate of overoptimistic results, e.g.---depending on the context---false positive test results and inflation of effect sizes \citep{simmons2011false,wasserstein2016asa,ioannidis2005most}, or, beyond the context of testing and effect estimation, to exaggerated measures of predictive performance \citep{boulesteix2009optimal} or clustering validity \citep{ullmann2023over}. 

\cite{hoffmann2021multiplicity} outline six sources of uncertainty that are omnipresent in empirical sciences and lead to variability of results in empirical research regardless of the considered discipline, namely sampling, measurement, model, parameter, data pre-processing, and method uncertainty. Failure to take these various uncertainties into account may lead to unstable, supposedly precise, but overoptimistic and thus potentially unreplicable results. Most importantly, model, parameter, data preprocessing and method uncertainties lead to the analytical variability mentioned above. 
In this context, \cite{simmons2011false} denote the flexibility researchers have regarding the different aspects of the analysis strategy as \lq\lq researcher degrees of freedom''. 

While it is clear that selective reporting of the \lq\lq most favorable results'' out of a multitude of results is a questionable research practice that invalidates statistical inference, it is less clear how researchers should deal with their degrees of freedom in practice. 
In this study, we suggest to tackle this issue from the perspective of multiple testing. More precisely, for analyses based on hypothesis testing we formalize researcher degrees of freedom as a multiple testing problem. We further propose to use an adjustment procedure to correct for the over-optimism resulting from the selection of the lowest p-value out of a variety of analysis strategies.

As the results of different analysis strategies addressing the same research question with the same data are usually highly dependent, a naive approach such as the Bonferroni correction is inappropriate. It would indeed lead to an unacceptable loss of power. Instead, we propose resorting to the single-step \lq\lq minP'' adjustment method \citep{westall1993multiplicity, westfall1993resampling} and discuss its use in this context. The  power achieved by the minP procedure is typically larger than with simpler approaches while ensuring a weak control of the family-wise error rate. This is because the procedure is based on the distribution of the minimal p-value, which is obviously affected by the level of  correlation between the tests.

The minP procedure has the major advantage that it has a relatively intuitive principle, as illustrated by the following example. In a comment on a study by \cite{mathews2008you} claiming that breakfast cereal intake before pregnancy is positively associated with the probability to conceive a male fetus, \cite{young2009cereal} reinterpret the small p-value of 0.0034 obtained in the original article. They notice that \cite{mathews2008you} did not only analyse the association between fetal sex and the consumption of breakfast cereals, but also many other food items---a typical case of multiple testing.  Based on the analysis of permuted data (i.e. data with randomly shuffled fetal sex status), \cite{young2009cereal} argue that \lq\lq one would expect to see a p-value as small as 0.0034 approximately 28 percent of the time when nothing is going on''. Implicitly, they apply the minP procedure for adjusting the smallest raw p-value of 0.0034 to 0.28 in this context where multiple tests are performed to investigate multiple food items. Our suggestion consists of translating this approach into the context of the analytical researcher degrees of freedom towards addressing the statistical factors of the replication crisis. 

The minP procedure as used in the example by \cite{young2009cereal} and considered in this paper is based on an approximation of the null distribution of the minimal p-value through a permutation-based procedure. We note, however, that such a permutation-based procedure is not always possible, and that resorting to theoretical asymptotical results on the distribution of the minimal p-value (or maximimal statistic) is more appropriate in some cases, as will be discussed later.

The goal of this paper can be seen as building bridges between two scientific communities. On one hand, the metascientific community has long recognized that the replication crisis in science is partly related to multiplicity issues, but has to date neither formalized the issue in terms of multiple testing nor applied known adjustment procedures for reducing the occurrence of false positive results. On the other hand, the multiple testing community is increasingly developing theoretically founded general approaches to multiple testing taking into account the dependence of the tests; see \cite{ristl2020simultaneous} for a recent important milestone. These approaches are however not yet routinely used to adjust for researcher degrees of freedom in practice. The reasons are manifold. The lack of communication between the two communities and the methodological complexity of these methods certainly play an important role. Another reason is that these approaches, even if increasingly efficient and general, do not address all types of analyses but only regression models, and require assumptions regarding the data format that may not always be fulfilled in practice. In this context, the present paper aims to formalize and demonstrate the use of minP to adjust for researcher degrees of freedom in simple situations not only involving linear models, while hopefully creating a common basis fostering communication between the two communities towards the development (by statistical researchers) and routine use (by applied data analysts) of more complex approaches. This paper aims to establish an easy approach designed to prevent the detection of false-positive findings in the context of fishing expeditions. 

The rest of this paper is structured as follows. Problems related to researcher degrees of freedom are outlined in more detail in Section~\ref{sec:background}, including potential approaches for handling it in practice that were proposed in the literature. As a motivating example, Section~\ref{sec:motivating_example} presents a study on the impact of perioperative partial arterial pressure ($paO_2$) on post-operative complications after neurosurgery that uses routinely collected real-world data. Our suggested approach is described in Section~\ref{sec:method}, while Section~\ref{sec:illustration} shows its results on the example dataset and Section~\ref{sec:discussion} briefly discusses limitations of the approach and possible extensions.

\section{Background: researcher degrees of freedom}
\label{sec:background}
\subsection{Overview}
\label{subsec:overview}
When analysing biomedical data, researchers are often confronted with a number of decisions that may appear trivial at first view, but  often have a considerable impact on study results. Which confounders should we adjust for? How should we handle missing values and outliers? Should we log-transform a continuous variable? What about categorical variables with categories that include no more than a handful of patients? Should these small categories be merged? Is a parametric or non-parametric test more appropriate? 
 The term \lq\lq researcher degrees of freedom'' \citep{simmons2011false} denotes, in a broad sense, this flexibility arising from the many analytical choices researchers face when analysing data in practice. 
 
 In most cases, neither theory nor precise practical guidance from the literature can reliably point researchers  to the \lq\lq best way'' to analyse their data. Model selection techniques based, e.g., on the Akaike Information Criterion (AIC) and diagnostic tools (e.g., to assess whether a variable is normally distributed) may be helpful in some cases. However, they most often do not provide definitive clear-cut answers to all the arising questions. Furthermore, the choice of these techniques is itself affected by uncertainty: there usually exist several suitable variants of them. For example, should we prefer the AIC or the Bayesian Information Criterion (BIC) for model selection? Should we use a QQ-plot or apply a test (if yes, which one and at which level?) to assess normality of a variable?

Combined with selective reporting, researcher degrees of freedom can lead to an increased rate of false positive results, inflation in effect sizes, and overoptimistic results \citep{ioannidis2005most, simmons2011false, wasserstein2016asa,hoffmann2021multiplicity}. The terms \lq\lq p-hacking'' and \lq\lq fishing for significance'' have been used in the context of hypothesis testing to denote the selective reporting of the most significant results out of a multitude of results arising through the multiplicity of analysis strategies. The resulting optimism is however not limited to the context of hypothesis testing. \lq\lq Fishing expeditions'' (also termed \lq\lq cherry-picking'' or \lq\lq data dredging'') are common issues in all types of analyses beyond hypothesis testing \citep{ullmann2023over}.

The multiplicity of possible analysis strategies particularly affects studies involving electronic health records and administrative claims data, which currently raise hopes and promises of ``real-world'' evidence and personalized treatment regimes. With data that have not been primarily collected for research purposes, uncertainties related to the analysis strategies may indeed be even more pronounced compared to the analysis of classical observational research data. In the last few years, contradictory results have been published in this setting, which can be viewed as a consequence of the uncertainties in a broad sense. See for example the conflicting results by \cite{fields2019does, childers2019re, childers2021same, turner2019utilization} on infectious complications associated with laparoscopic appendectomies and those by \cite{jivanji2020association, shah2021association} on the association between cardiovascular disease and marijuana consumption. In both cases, different teams of researchers used the same data set to answer the same research question and found contradictory results which can be explained by seemingly trivial choices. 

\subsection{Partial solutions and related work}
\label{subsec:related_work}
There are a number of approaches that have been proposed to deal with uncertainty regarding the analysis strategy and are preferable to the selective reporting of the preferred results. 

A natural approach is to  fix the analysis strategy in advance, i.e. prior to running the analyses, to avoid obtaining multiple results in the first place. For more transparency, this may be done within a publicly available pre-registration document \citep{nosek2018preregistration, munafo2017manifesto, hardwicke2023reducing}, thus preventing result-dependent selective reporting \citep{naudet2024}. This type of pre-registration is the standard for clinical trials \citep{chan2013spirit}. However, even in the strictly regulated context of clinical trials, there is some controversy about the question whether statistical analysis plans of clinical trials are detailed enough \citep{greenberg2018pre} to prevent potential selective reporting. Fixing the analysis strategy in advance  tends to be even more difficult for exploratory research questions and for complex data sets and research questions. 

The opposite approach consists of transparently acknowledging uncertainty and reporting the variety of results obtained with the considered analysis strategies.  This concept has been proposed in different variants in the last decade: it encompasses, e.g., the vibration of effect framework \citep{patel2015assessment,klau2023comparing}, multiverse analyses \citep{steegen2016increasing} and the specification curve analysis \citep{rohrer2017probing,simonsohn2020specification}.
With these approaches, the multiple reported results might be conflicting, sometimes yielding a confusing picture and a paper without clear-cut take-home message. In other words, the pitfalls of selective reporting are obviously avoided, but this comes at a high price in terms of interpretability and clarity.

Finally, let us mention the approach of conducting various analyses, selecting the preferred results but---instead of reporting it in a cherry-picking fashion---publishing it only if it can be qualitatively confirmed by running the exact same analysis on independent \lq\lq validation'' data \citep{daumer2008reducing}. This is the approach \cite{ioannidis2005microarrays} indirectly recommends when claiming {\it \lq\lq Without highly specified a priori hypotheses, there are hundreds of ways to analyse the dullest dataset. Thus, no matter what my discovery eventually is, it should not be taken seriously,  unless it can be shown that the same exact mode of analysis gets similar results in a different dataset.''}
This approach, however, requires to set apart (or subsequently obtain) a validation dataset of adequate size. This might not always be possible, and even in cases where it is possible, splitting the data may imply a substantial loss of power compared to the analyses that would have been performed using the totality of the data \citep{daumer2008reducing}.

In the context of analyses strongly affected by uncertainties where none of these simple approaches seems applicable, we suggest an alternative approach based on multiple testing correction. More specifically, we view researcher degrees of freedom from a multiple testing perspective and propose to apply correction for multiple testing to the preferred result to reduce the risk of type 1 error, as outlined in Sections~\ref{subsec:RDM_MT} and \ref{subsec:FWER}.

\section{Motivating example}
\label{sec:motivating_example}
\subsection{Data}
\label{subsec:data}
As a motivating example, we use a current research project on the effect of partial arterial pressure of oxygen ($paO2$) during craniotomy on post-operative complications among neurosurgical patients. This study is based on a routinely collected dataset from a Munich University Hospital preprocessed as described in \cite{Becker-Pennrich2022}. 

While the irreversible damage to the brain caused by reduced levels of oxygen in the blood (hypoxemia) has been the topic of extensive research, the potential harm caused by an increased amount of oxygen (hyperoxemia) is comparatively not well understood. The dangers of over-supplementation of oxygen during surgical procedures are still debated among anesthesiologists and a topic of current research \citep{mcilroy2022oxygen, weenink2020perioperative}. 

The dataset under consideration was extracted from routine clinical care data of $n=3,163$ surgical procedures performed on lung healthy neurosurgical patients. Vital data was measured at several timepoints during surgery for each surgical procedure. As outlined in \cite{Becker-Pennrich2022}, measuring $paO2$ continuously is not feasible, in contrast to other vital parameters. To obtain a reliable assessment of hyperoxemia during the surgical procedure, the $paO2$ values thus have to be imputed using a surrogate model based on proxy variables that can be measured continuously using non-invasive techniques. \cite{Becker-Pennrich2022} suggest to use machine learning methods for this purpose and identify random forest, and regularized linear regression as well-performing candidates.

In this paper, we consider the assessment of the effect of $paO2$ on the binary outcome defined as the occurrence of post-operative complications after surgery. Even if we ignore model choice issues arising from the selection of a set of potential confounders, this analysis is characterized by a large number of uncertain choices. They are described in more detail in Section~\ref{subsec:paO2_RDF} along with the options considered in our illustrative study in Section~\ref{sec:illustration}.

\subsection{Researcher degrees of freedom}
\label{subsec:paO2_RDF}
\begin{figure}
\begin{center}
\begin{tikzpicture}[scale=0.9]
\node {Raw data} [sibling distance = 3.5cm, dashed]
    child [teal, align = center] {  node {Drop missings \\ \vdots}  }
    child [teal] {  node  {Impute missings}
                    child [teal, align = center] { node {Random forest imputation \\ \vdots} }
                    child [teal] { node {GLMnet imputation} 
                            child [teal, align = center] { node {Parameter tuning \\ \vdots} }
                            child [teal] { node {No parameter tuning} 
                                    child [teal, align = center] { node {Mean aggregation \\ \vdots} }
                                    child [teal] { node {Median aggregation}
                                        child [brown] { node {Logistic regression} }
                                        child [brown] { node {Fisher's test} }
                                        }
                                    }
                            }
             }  ;
\end{tikzpicture}
\caption{Overview of the different researcher degrees of freedom. All in all 48 specifications were analyzed. Green depicts the data pre-processing decisions while brown depicts the method choices.}
\end{center}
\label{fig:specs}
\end{figure}
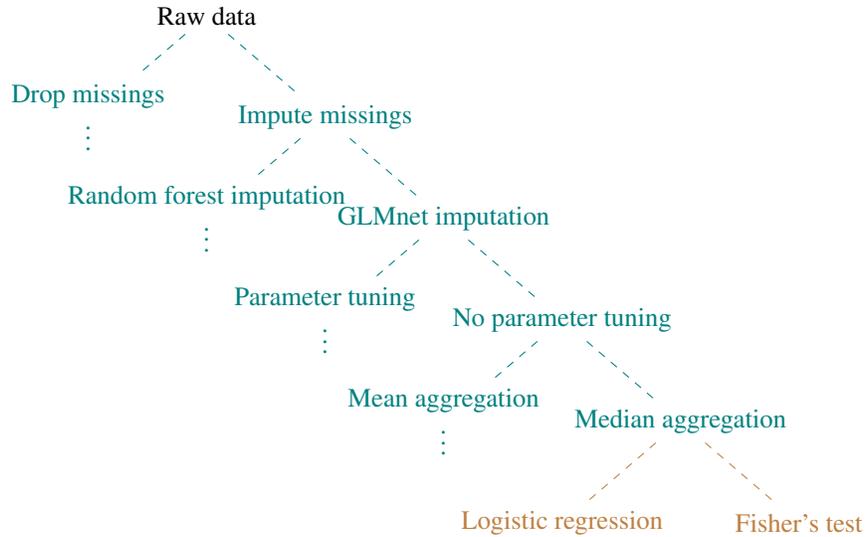

 In our study, we  focus on the following choices, depicted in the form of a decision tree in Figure~\ref{fig:specs}: (i) missing value imputation, (ii) surrogate model for the unobserved $paO2$-values, (iii) parameter choice approach, (iv) aggregation procedure, and (v) coding of the exposure variable $paO2$ and testing method. Uncertainty (ii) is discussed in more details by \cite{Becker-Pennrich2022}. In this study, we use the data preprocessed as described in \cite{Becker-Pennrich2022} resulting from the different surrogate modelling strategies.

Uncertainties (i) to (iv) can be seen as {\it preprocessing uncertainty} in the terminology of \cite{hoffmann2021multiplicity}. For the missing value imputation (i) the two considered options are to either drop or impute the missing values using multiple imputation in the 'mice' package \citep{mice}. For surrogate modelling  of the unobserved $paO2$-values (ii) we either use random forest or a regularized general linear model, either using the default parameter values or the parameter values obtained through tuning via random search using predefined tuning spaces (iii) as implemented in the 'mlr3' package \citep{mlr3}. 

After obtaining a prediction of unobserved $paO2$ values through surrogate modelling, for each surgery the $paO2$ measurements are aggregated to a single value over multiple measurements for a single patient: either the mean or the median (iv). Finally (v), we either consider $paO2$ as a continuous variable and use a logistic regression model to assess its effect on the binary outcome, we dichotomize it using the clinically meaningful cutoff value of 200mmHg, or we categorize it into a three-category variable using the clinically meaningful cutoff values of 200mmHg and 250mmHg and use Fisher's exact test. The latter choice can be seen as referring both to preprocessing and method uncertainty, since the choice of the test is related to the transformation of the variable $paO2$. 

All in all, we consider a total of 48 specifications of the analysis strategy:  2 (missing values) $\times$ 2 (surrogate model) $\times$ 2 (parameter choice) $\times$ 2 (aggregation) $\times$ 3 (method) = 48.

\section{Method}
\label{sec:method}

\subsection{Researcher degrees of freedom as a multiple testing problem}
\label{subsec:RDM_MT}
In the remainder of this paper, we will focus on analyses that consist of statistical tests. We consider a researcher investigating a---possibly vaguely defined---research hypothesis such as \lq\lq {\it $paO2$} has an impact on post-operative complications'', as opposed to the null- and alternative hypotheses of a formal statistical test, which are precisely formulated in mathematical terms. From now on, we assume that the research hypothesis the researcher wants to establish corresponds to the formal alternative hypothesis of the performed tests.

In this context, the term \lq\lq analysis strategy'' refers to all steps performed prior to applying the statistical test as well as to the features of the test itself. The following aspects can be seen as referring to {\it preprocessing uncertainty} in the terminology by \cite{hoffmann2021multiplicity}: transformation of continuous variables, handling of outliers and missing values, or merging of categories. Aspects related to the test itself refer to {\it model and method uncertainty} in the terminology of \cite{hoffmann2021multiplicity}. They include, for example, the statistical model underlying the test, the formal hypothesis under consideration, or the test (variant) used to test this null-hypothesis.

In the context of testing, an {\it analysis strategy} can be viewed as a combination of such choices. Obviously, different analysis strategies will likely yield different p-values and possibly  different test decision (reject the null-hypothesis or not). Applying different analysis strategies successively to address the same research question amounts to performing multiple tests. 
From now on, we denote $m$ as the number of analysis strategies considered by a researcher. The null-hypotheses tested through each of the $m$ analyses are denoted as $H_{0}^{i}$,\ $i=1,\dots,m$.

These null-hypotheses and the associated alternative hypotheses can be seen as---possibly different---mathematical formalizations of the vaguely defined research hypothesis---\lq\lq $paO2$ has an impact on post-operative complications'' in our example. One may decide to formalize this research hypothesis  as \lq\lq $H_0:$\ the mean $paO2$ is equal in the groups with and without post-operative complications versus $H_1:$\ the mean $paO2$ is not equal in these two groups''. But it would also be possible to formalize it as \lq\lq $H_0$: the post-operative complication rates are equal for patients with $paO2<200$mmHg and those with $paO2\geq 200$mmHg'' versus \lq\lq $H_1:$\ the post-operative complication rates are not equal for patients with $paO2<200$mmHg and those with $paO2\geq 200$mmHg''. Analysis strategies may thus differ in the exact definition of the considered null- and alternative hypotheses.

They may, however, also differ in other aspects, some of which were mentioned above (for example the handling of missing values or outliers). If two analysis strategies $i_1$ and $i_2$ (with $1\leq i_1<i_2\leq m$) consider exactly the same null-hypothesis, we  have $H_0^{i_1}=H_0^{i_2}$. Of course, it may also happen that the research hypothesis is not vaguely defined but already formulated mathematically as null- and alternative hypotheses, and that the $m$ analysis strategies thus only differ in other aspects such as the handling of missing values or outliers. In this case the $m$ null-hypotheses would all be identical.

Regardless whether the hypotheses $H_0^{i}$ ($i=1,\dots,m$) are (partly) distinct or all identical, a typical researcher who exploits the degree of freedom by \lq\lq fishing for significance'' performs the $m$ testing analyses successively. They hope that at least one of them will yield a significant result, i.e. that the smallest  p-value, denoted as $p_{(1)}$, is smaller than the significance level $\alpha$. If it is, they typically report it as convincing evidence in favor of their vaguely defined research hypothesis. It must be noted that in this hypothetical setting the researcher is not interested in identifying the \lq\lq best'' model or analysis strategy but only in reporting the lowest p-value that supports the hypothesis at hand. 

Considering this scenario from the perspective of multiple testing, it is clear that the probability to thereby make at least one type 1 error, denoted as Family Wise Error Rate (FWER), is possibly strongly inflated. In particular, even if all tested null-hypotheses are true, we have a  probability greater than $\alpha$ that the smallest p-value $p_{(1)}$ is smaller than $\alpha$; this is precisely the result researchers engaged in fishing for significance will report. This problem can be seen as one of the explanations as to why the proportion of false positive test results among published results is substantially larger than the considered nominal significance level of the performed tests \citep{ioannidis2005most}.

\subsection{Controlling the Family-Wise Error Rate (FWER)}
\label{subsec:FWER}
Following the formalization of researcher degrees of freedom as a multiple testing situation, we now consider the problem of adjusting for multiple testing in order to control the FWER. 
More precisely, we want to control the probability $P(\text{Reject at least one true}\, H_0^{i}) $ to make at least one type 1 error when testing $H_0^{1},\dots,H_0^{m}$, i.e. the FWER. 

More precisely, we primarily want to control the FWER in case all null-hypotheses are true. Imagine a case where some of the null-hypotheses are false and there is at least one false positive result. On one hand, if $p_{(1)}$ is not among the falsely significant p-values, the false positive test result(s) typically do(es) not affect the results ultimately reported by the researchers (who focus on $p_{(1)}$). This situation is not problematic. On the other hand, if $p_{(1)}$ is falsely significant, $H_0^{(1)}$ is {\it wrongly} rejected, and strictly speaking a false positive result (\lq\lq $p_{(1)}<\alpha$'') is reported.  However, some of the $m-1$ remaining null-hypotheses, which are closely related to $H_0^{(1)}$ (because they formalize the same vaguely defined research hypothesis), {\it are} false. Thus, rejecting $H_0^{(1)}$ is not fundamentally misleading in terms of the vaguely defined research hypothesis. As a result, in the context of the researcher degrees of freedom, false positives have to be avoided primarily in the case when all null-hypotheses are true.

In other words, we need to control the probability $P(\text{Reject at least one true}\, H_0^{i} | \cap_{i=1}^mH_0^{i})$ to have at least one false positive result {\it given} that all null-hypotheses are true, i.e. we want to achieve a weak control of the FWER.
Various adjustment procedures exist to achieve strong or weak control of the FWER; see \cite{dudoit2003multiple} for concise definitions of the most usual ones (including those mentioned in this section).

The most well-known and simple procedure is certainly the Bonferroni procedure. It achieves strong control of the FWER, i.e. it controls $P(\text{Reject at least one true}\, H_0^{i})$  under any combination of true and false null hypotheses. This procedure adjusts the significance level to $\tilde{\alpha} = \alpha/m$; or equivalently it adjusts the p-values $p_i$ ($i=1,\dots,m$) to $\tilde{p_i} = \min(mp_i,1)$. However, the Bonferroni procedure is known to yield low power in rejecting wrong null-hypotheses in the case of strong dependence between the tests. The so-called Holm stepwise procedure, which is directly derived from the Bonferroni procedure, has a better power. However, the Holm procedure adjusts the smallest p-value $p_{(1)}$ exactly to the same value as the Bonferroni procedure. It implies that, if none of the $m$ tests lead to rejection with the Bonferroni procedure, it will also be the case with the Holm procedure. The latter can thus not be seen as an improvement over Bonferroni in terms of power in our context, where the focus is on the smallest p-value $p_{(1)}$.

\subsection{The minP-procedure}
\label{sec:minP_procedure}
The permutation-based minP adjustment procedure for multiple testing \citep{westall1993multiplicity} indirectly takes the dependence between tests into account by considering the distribution of the {\it minimal} p-value out of $p_1,\dots,p_m$. This increases its power in situations with high dependencies between the tests, and thus makes it a suitable adjustment procedure to be applied in the present context. In the general case it controls the FWER only weakly, but as outlined above we do not view this as a drawback in the present context.

The rest of this section briefly describes the single-step minP adjustment procedure based on the review article by \cite{dudoit2003multiple}.
The following description is not specific to researcher degrees of freedom considered in this paper. However, for simplicity we further use the notations ($p_i$, $H_0^i$, for $i=1,\dots,m$) already introduced in Section~\ref{subsec:RDM_MT} in this context.

In the single-step minP procedure, the adjusted p-values $\tilde{p}_i$, $i=1,\dots,m$ are defined as
    \begin{equation}
    \Tilde{p}_i = P \left( \min_{1 \leq \ell \leq m} P_\ell \leq p_i \mid \cap_{i=1}^mH_0^{i}\right),
    \label{eq:minP}
\end{equation}
with   $P_\ell$ being the random variable for the unadjusted p-value for the $\ell^{th}$ null-hypothesis $H_0^\ell$ \citep{dudoit2003multiple}. The adjusted p-values are thus defined based on the distribution of the minimal p-value out of $p_1,\dots,p_m$, hence the term \lq\lq minP''. In the context of the researcher degrees of freedom considered here, the focus is naturally on $\tilde{p}_{(1)}= P \left( \min_{1 \leq \ell \leq m} P_\ell \leq p_1 \mid \cap_{i=1}^mH_0^{i}\right)$.

In many practical situations, including the one  considered in this paper, the distribution of $\min_{1 \leq \ell \leq m} P_\ell$ is unknown. The probability in Eq. (\ref{eq:minP}) thus has to be approximated using permuted versions of the data that mimic the global null-hypothesis $\cap_{i=1}^mH_0^{i}$. More precisely, the adjusted p-value $\tilde{p}_i$ is approximated as the proportion of permutations for which the minimal p-value is lower or equal to the p-value $p_i$ observed in the original data set. Obviously, the number of permutations has to be large for this proportion to be estimated precisely.
In the example described in Section~\ref{sec:motivating_example} involving only two variables ($paO2$ and post-operative complications), permuted data sets are simply obtained by randomly shuffling one of the variables. More complex cases will be discussed in Section~\ref{sec:discussion}.

\section{Illustration}
\label{sec:illustration}

\subsection{Study design}
\label{subsec:study_design}
The study aims at illustrating the use and behavior of the minP-based approach when used to adjust for the multiplicity arising through researcher degrees of freedom. We use the original as well as permuted versions of the $paO2$ data set. The 48 specifications of the analysis strategy outlined in Section~\ref{sec:motivating_example} are successively applied. P-values are either left unadjusted, or adjusted using the Bonferroni procedure, or adjusted using the recommended minP procedure with 1000 permutations. 
All analyses are performed for different sample sizes. Subsets of each considered size are randomly drawn from the original data set without replacement. 

The study consists of two distinct parts. In the first part, we assess the family-wise error rate (FWER) for different sample sizes with the three approaches (no adjustment, Bonferroni adjustment, and minP adjustment).  For this purpose, we generate data  without  association between the two variables of interest ($paO2$ and the outcome \lq\lq post-operative complications'') by using a $paO2$ covariate vector drawn without replacement from the true dataset but randomly generating the binary outcome variable from a binomial distribution ($p=0.5$) to break the association between the outcome and $paO2$. This procedure is repeated 1000 times for every $n \in \{100,200,300,500,2000,3000\}$. For each run, we calculate unadjusted, minP-adjusted, and Bonferroni-adjusted p-values as outlined above and check whether there is at least one false positive, i.e. whether at least one of the respective p-values of the 48 specification is significant at the 5\% level. The proportion of the 1000 runs for which this happens yields an estimate of the FWER of the three approaches.

In the second part, the original data set is analysed. Based on medical knowledge we expect a strong relationship between $paO2$ and the outcome to be present, but do not formally know the truth. 
For each of the three approaches (no adjustment, Bonferroni adjustment, and minP adjustment), we calculate the proportion of significant p-values at the 1\%, 5\% and 10\% level among the 48 specifications. This was repeated 1000 times for each sample size $n \in (50,100,150,200,250,300)$. As in our example study, the association becomes highly significant for larger sample sizes and all p-values are then very close to zero, we only focus on these small sample sizes here. The code for reproducing the analyses can be found on GitHub\footnote{\url{https://github.com/mmax-code/researcher_dof}}.

\subsection{Results}
\label{subsec:results}
\begin{figure}[ht]
    \centering
    \includegraphics[scale = 0.3]{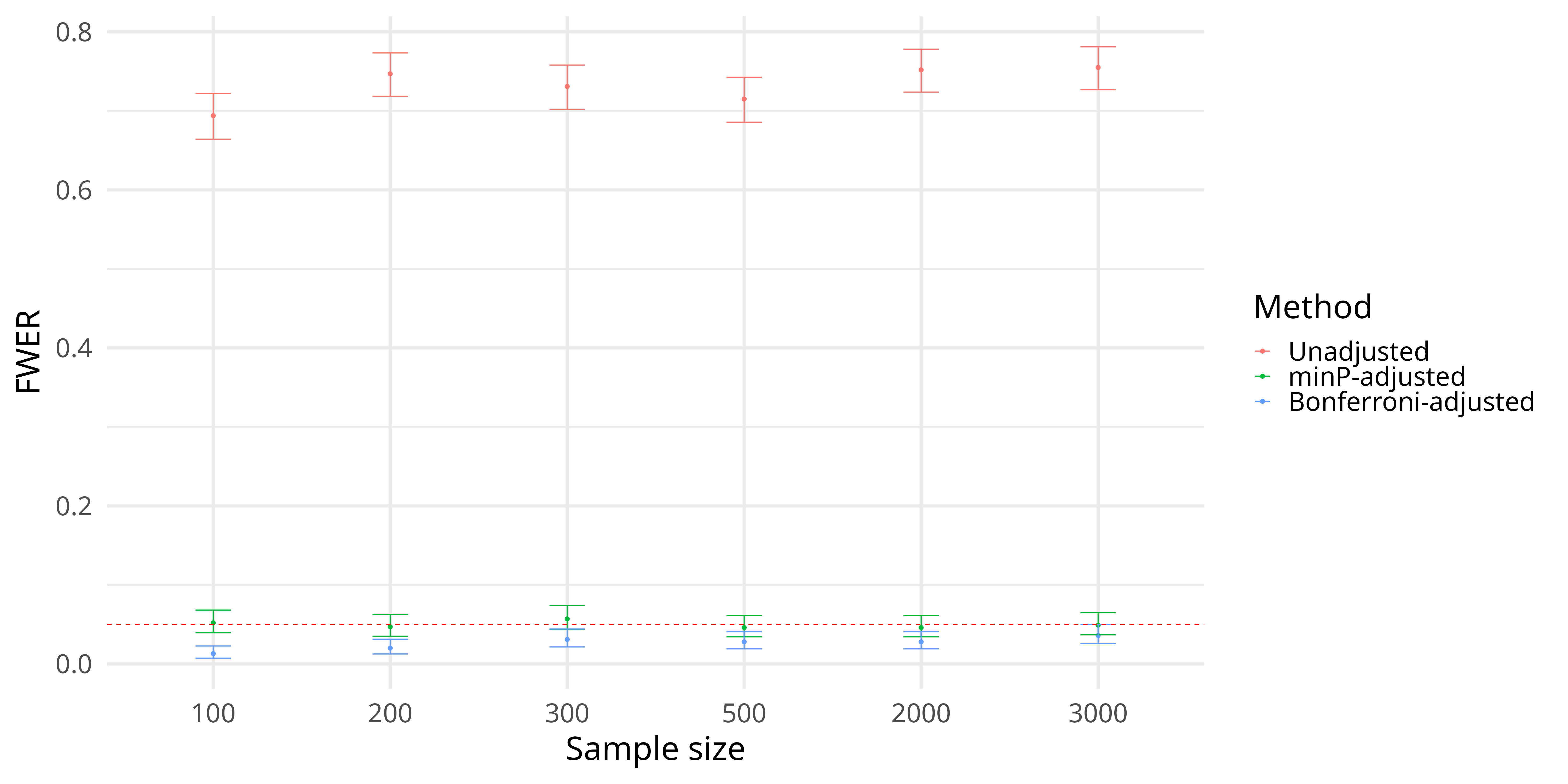}
    \caption{FWER with Newcombe confidence intervals (computed over 1000 simulation runs) for different sample sizes without an association between post-operative complications and $paO2$. Dashed red line indicates 5\% significance level. }
    \label{fig:fwer}
\end{figure}

Figure~\ref{fig:fwer} shows the estimated FWER for different sample sizes along with the Newcombe confidence intervals \citep{newcombe1998interval}. In the absence of adjustment, false-positive results appear to be present in at least one of the 48 specifications for about 70\% of the data sets of size $n=100$ and 76\% of the data sets of size $n=3000$, which aligns with the results of \citep{simonsohn2020specification}. If we adjust the p-values using the minP-approach (green), the 5\% level is held for all considered sample sizes. As expected the Bonferroni adjustment (blue) is more conservative: the confidence intervals for FWER, which do not include 0.05,  only overlap with those of the minP procedure for a sample size of $n=3000$. 

\begin{figure}[ht]
    \centering
    \includegraphics[scale = 0.3]{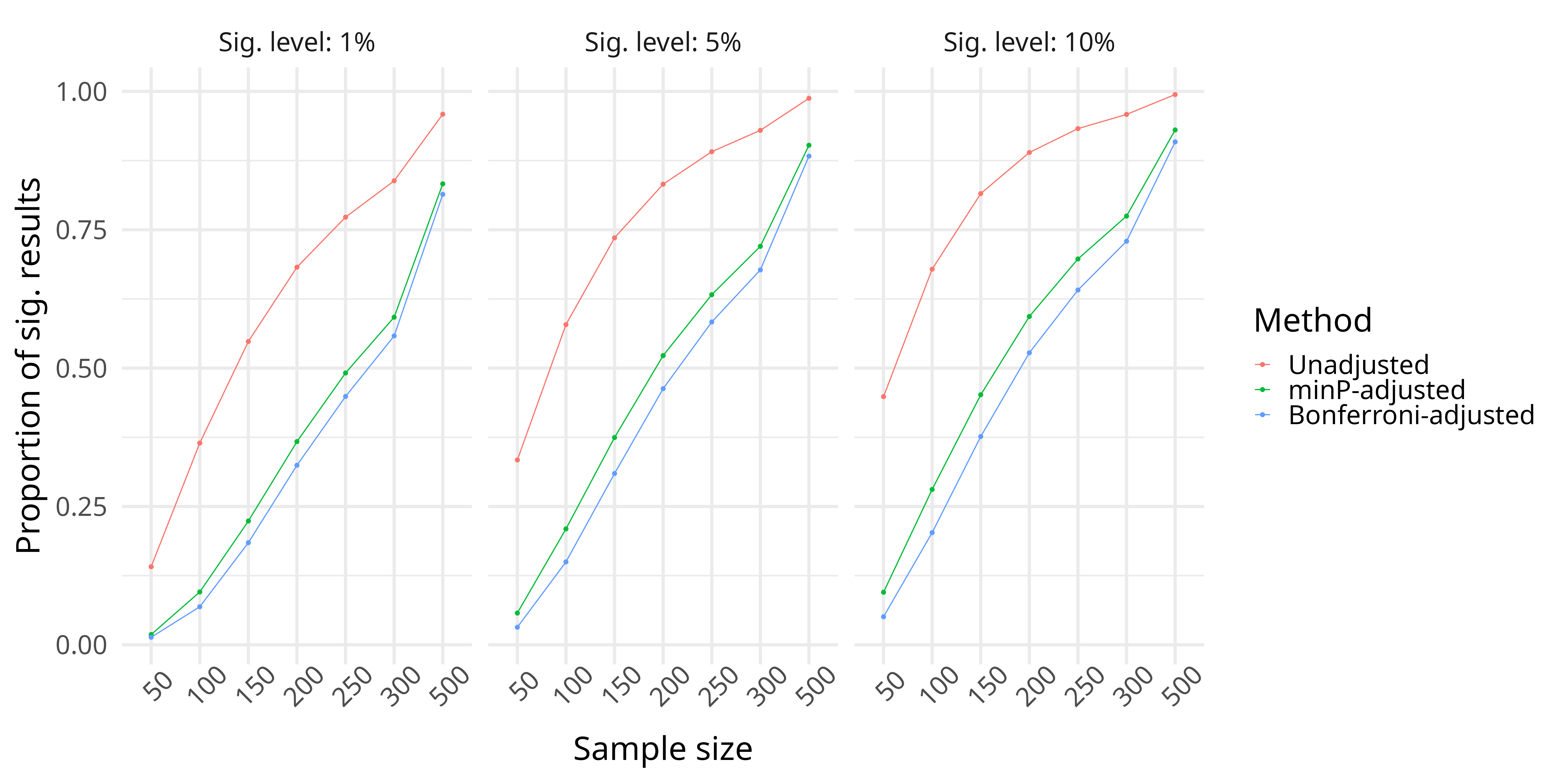}
    \caption{Proportion of significant results for all 48 specifications for $\alpha \in (0.01,0.05,0.1)$ and sample size $n \in (50,100,150,200,250,300,500)$. Line colors indicate results based on unadjusted (red), minP-adjusted (green) and Bonferroni-adjusted (blue) p-values.}
    \label{fig:prop}
\end{figure}

Figure~\ref{fig:prop} presents the proportion of significant p-values at the 1\%, 5\% and 10\% level over the 48 specifications for the three approaches and different sample sizes. These proportions are averaged over 1000 runs. As we expect a highly significant association between the two variables of interest, we focus on small sample sizes only. The observed trend is not surprising: For all $n \in (50,100,150,200,250,300,500)$ it holds that
\[
\overline{\sum_{i=1}^{48}\mathbf{1}(p_{i_{\text{{unadjusted}}}}<\alpha)/48} > 
\overline{\sum_{i=1}^{48}\mathbf{1}(p_{i_{\text{{minP}}}}<\alpha)/48} > 
\overline{\sum_{i=1}^{48}\mathbf{1}(p_{i_{\text{{bonferroni}}}}<\alpha)/48},
\]
 (where the overline stands for the average over 1000 runs and $\alpha \in (0.01, 0.05, 0.1)$), i.e. more significant results appear for the unadjusted p-values compared to the adjusted p-values. Furthermore, the Bonferroni approach is more conservative than the minP-adjustment.

\section{Discussion}
\label{sec:discussion}
In this work, we described a framework for performing valid statistical inference in the presence of researcher degrees of freedom  through adjustment for multiple testing. Our results on simulated data and in an application concerning $paO2$ and post-operative complications suggest that the minP procedure is appropriate for this purpose. 

The use of permutation-based procedures has already been recommended by \cite{simonsohn2020specification} to address researcher degrees of freedom. There are, however, fundamental differences between this approach and ours. \cite{simonsohn2020specification} address the problem of researcher degrees of freedom by specifying all plausible specifications (analysis strategies in our terminology) and ultimately evaluating the joint distribution of the estimated effects of interest across these model specifications. This evaluation is done graphically through the so-called specification curve, but also through a permutation test addressing null-hypotheses such as \lq\lq the median effect across the specifications is zero''. 

This approach, while  similar to ours at first view and interesting, is different in several aspects. Firstly, permutations are used by \cite{simonsohn2020specification} as part of a permutation-based test and not within a multiple testing adjustment procedure. Our suggestion is precisely to formalize the multiplicity of analysis strategies as a multiple testing problem---and to benefit from various methodological results obtained in the field, for example on the weak control of the FWER through the minP procedure.  That said, minP adjustment can be viewed as a simple permutation test for the test statistic \lq\lq minimal p-value'', hence the apparent similarity with the permutation test for the median effect.

Secondly, and more importantly, the focus on the {\it median effect} makes the procedure by \cite{simonsohn2020specification} sensitive to misspecifications that do not model the data properly and thus fail to show an effect even if there is one. Imagine a fictive example where one runs 99 fully inappropriate analyses yielding non-significant results and one meaningful analysis that identifies a highly significant (truly existing) effect. The true median effect is zero, and the permutation test by \cite{simonsohn2020specification} will certainly not reject the null. In contrast, with our approach the truly existing effect is likely to be detected by the meaningful analysis. This is because the minP procedure focuses on the {\it minimal} p-value, which is very small in this fictive example. This focus on the minimal p-value better accounts for the fact that, in practice, one would often include some analysis strategies that are in fact inappropriate to detect the effect of interest. It also better reflects the common p-hacking practice that consists of selecting and reporting the smallest p-value. However, our approach raises a number of questions that may be addressed in future research.

Firstly, the specification of an appropriate permutation procedure taking the data and the specificity of the research question into account is not always easy/possible. Let us consider the following example: the null-hypothesis of interest is that the means of a variable are equal in two groups, while the variances may be different in the two groups. By permuting the group labels, one also inevitably enforces equality of the variances, which is a stronger assumption than the null-hypothesis of interest \citep{dudoit2003multiple}.
Defining a permutation scheme that reflects the global null-hypothesis $\cap_{i=1}^mH_0^{i}$ may also be intricate in the case of multivariable regression models involving confounders in addition to the exposure of interest whose effect on the dependent variable is to be investigated. On the one hand, permuting only the exposure of interest will destroy the association between this exposure and confounders. On the other hand, permuting the outcome will not only destroy the association between exposure variable and dependent variable, but also the association between the confounders and the outcome. In principle, none of these simple permutation procedures are suitable. Both enforce more than the considered null-hypothesis of no effect of the exposure on the outcome. Complex alternative permutation procedures may be preferred \citep{berrett2020conditional}. Alternatively, if all analysis strategies are based on marginal generalized estimating equation models, one may resort to asymptotical results on the distribution of the maximally selected statistic to derive adjusted p-values, thus avoiding time-consuming and methodologically complex permutation procedures; see for example \cite{ristl2020simultaneous}. Even though this approach is extremely powerful for most cases, it comes at the cost of some assumptions that are not applicable in our case (restrictions regarding the input data and focus on parametric tests).

Secondly, it would be interesting to investigate the behavior of our suggested approach compared to the validation approach mentioned in Section~\ref{subsec:related_work}, that consists of splitting the data into two parts, applying all candidate analysis strategies to the first part, and validating the preferred result by applying the analysis strategy that was used to obtain it to the second part of the data. Both this splitting procedure and the adjustment for multiple testing suggested in this paper imply a loss of power compared to the unadjusted analysis one would perform with the selected analysis strategy on the whole dataset. Researchers may prefer to run analyses on the whole dataset without arbitrary splitting, which is a clear argument in favor of our adjustment approach. However, the concept of validation using independent data may also seem attractive. Preference for one or the other approach is a matter of perspective. But, the power resulting from these two approaches may yield a decisive argument in favor for one of them. 

Finally, note that our paper should not be understood as a plea for the use of p-values in general. We merely claim that, if statistical testing is used and several analysis variants are performed, it certainly makes sense to adjust for multiplicity before interpreting these p-values. Our approach allows to selectively report the results of the analysis strategy yielding the most convincing evidence, while controlling the type 1 error---and thus the risk of publishing false positive results that may not be replicable. In future research, it approach could in principle be extended beyond the context of hypothesis testing.

\section*{Acknowledgments}
We thank Savanna Ratky for valuable language corrections and F. Julian Lange and Ludwig Hothorn for helpful comments. The authors gratefully acknowledge the funding by DFG grants BO3139/7-1 and BO3139/9-1 to Anne-Laure Boulesteix.

\bibliography{references}

\end{document}